\documentclass[3p, twocolumn]{elsarticle}

\usepackage{hyperref}

\usepackage{geometry}
\usepackage{graphicx}
\usepackage{amsmath}
\usepackage{amssymb}

\usepackage{booktabs}

\usepackage[utf8]{inputenc} 

\usepackage[T1]{fontenc}
\usepackage{lmodern}

\usepackage[ruled, linesnumbered]{algorithm2e}
\usepackage{algpseudocode}

\usepackage{tabularx}

\usepackage{xcolor}
\def\HiLi{\leavevmode\rlap{\hbox to \hsize{\color{yellow!50}\leaders\hrule height .8\baselineskip depth .5ex\hfill}}}

\usepackage{fancyvrb}
\usepackage{listings}

\lstdefinelanguage{CAL} {keywords={actor, action, do, od, if, fi, end, schedule, guard, fsm, proctype, chan, of, int, uint, priority}}

\begin{document}

\begin{frontmatter}

\title{Online Fault Identification of Digital Hydraulic Valves\\ Using a Combined Model-Based and Data-Driven Approach}

\author[1]{Johan Ersfolk}
\author[2]{Miika Ahopelto}
\author[1]{Wictor Lund}
\author[1]{Jonatan Wiik}
\author[1]{\\Marina Waldén}
\author[2]{Matti Linjama}
\author[1]{Jan Westerholm}
\address[1]{Åbo Akademi University, Finland}
\address[2]{Tampere University of Technology, Finland}

\begin{abstract}
Robustness and fault-tolerance are desirable properties for hydraulic working machines and field robots. In applications where service personnel do not have easy access to the machine, it is important that the machine can continue its operation despite single machinery faults. Digital hydraulics enables a design which is inherently fault-tolerant by having each hydraulic actuator controlled by a number of parallel on/off valves. The exact state operation of a digital hydraulic system enables model based diagnostics of the hydraulic components and the possibility to compensate for identified faults. This paper presents an approach for identifying faulty valves based on combination of pressure measurements made during the normal operation of the machine and a mathematical model describing the flow balance of the hydraulic system. 
\end{abstract}


\end{frontmatter}


\section{Introduction}

Modern field robots and work machines are highly automated and are often used for tasks which require high precision. In some applications, such as operation in hazardous areas, a field robot is not only required to complete its tasks without the presence of an operator, but also to continue its operation despite single machinery faults. This need for fault-tolerance requires machines with redundant hardware, the ability to perform online diagnostics identifying the faulty machine parts, and a control system that can compensate for the faults. Digital hydraulics addresses this need for fault tolerance by using simple and robust components which have precise and known control states~\cite{Linjama:2011}. By controlling each hydraulic flow path by several parallel connected valves, a controller which is aware of a faulty valve can find an alternative control state that compensate for the fault.

\begin{figure}[t]
\centering
\includegraphics[width=0.65\linewidth, trim={0 5cm 20cm 0},clip]{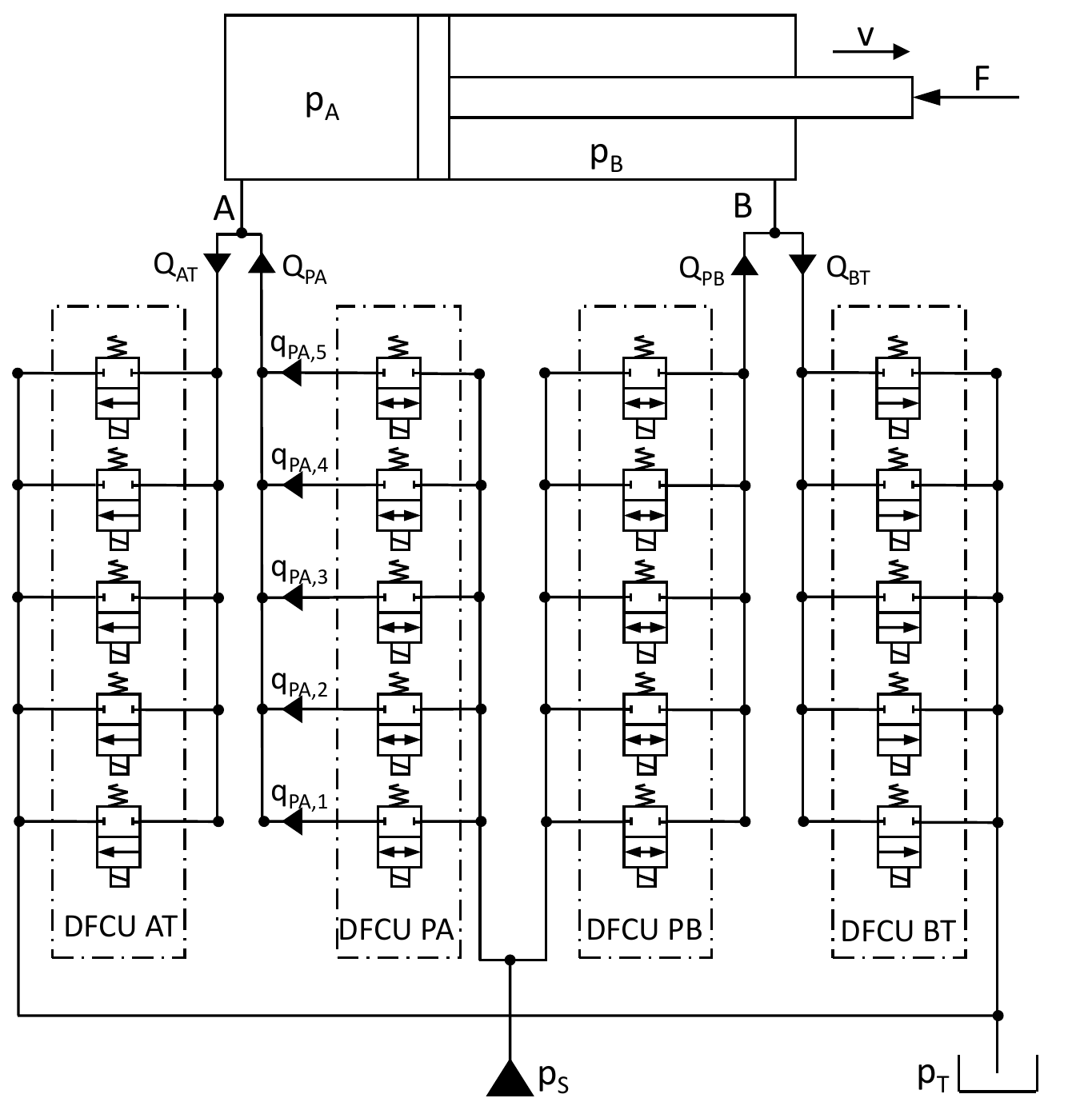}
\caption{A hydraulic actuator with four DFCUs each containing five binary valves. $p_A$ and $p_B$ are the pressures in the cylinder chambers while $p_S$ is the supply pressure and $p_T$ is the pressure in the tank.}
\label{fig:cylinder}
\end{figure}

A digital valve system uses a number of simple on/off-valves to achieve discrete controllability of hydraulic actuators. For each flow path, traditional valves are replaced with digital flow control units (DFCU) which consist of a number of parallel binary on/off-valves. When the flow capacities of the \emph{N} valves of a DFCU follow powers of two, the flow path can be controlled with $2^N$ unique control states. A typical system, like the one shown in Figure~\ref{fig:cylinder}, having four DFCUs with five valves each thus provides $2^{20}$ (1,048,576) unique control states. Efficient control of a digital hydraulic system can be achieved by evaluating and choosing control states that optimize the operation of the machine according to different criteria~\cite{jedfp16, linjama2007}. One jammed valve implies that the machine has to be controlled with the remaining $2^{19}$ control states.

The exact operation of digital valves makes it possible to control a digital hydraulic system with a model based controller which evaluates the steady state of a hydraulic actuator for several different valve states and chooses the best candidate. In a similar fashion, an exact model of the system makes it possible to observe the difference between the measured behavior of the machine and the expected behavior predicted by the steady state model. In this paper we show that it is possible to identify a broken valve and the position in which it is jammed, by simply analyzing the pressures in a cylinder's chambers and the supply pressure. 
This is an important feature as it enables diagnostics, of the digital valve system, without requiring any additional sensors, instead we only need the sensors that already are present in a system.

The approach presented in this paper operates online, during the normal operation of a hydraulic machine, by passively analyzing the pressure measurements and control signals without interfering with the machine operation. The approach is targeted towards analyzing the performance of a large number of valves while only sampling a relatively small number of pressure signals. 
In our example case we have 20 valves for which we consider single faults in both open and closed positions, thus we have 40 potential faults, while we only use three pressure sensors. The measurements from these few pressure sensors are collected over short periods (e.g. a hundred samples corresponding to one second) during various conditions and control states. This enables us to compute a likely cause for a deviation between the model and the measurements, which in turn enables the machine to quickly identify failing components and to automatically compensate for this without interrupting the current task. In case service is possible, our method provides information regarding which valve is malfunctioning.

\section{Background and Related Work}

In work machines, a component failure is both a safety and cost concern. For a modern work machine, with a large number of components, it can be expected that parts such as valves and hoses break. As it is not possible to completely avoid faults, a fault-tolerant design can be used to guarantee that the system is fully operational despite single machinery faults~\cite{blanke:2006}. For a digital hydraulic system to continue its operation despite broken valves, the faults need to be identified and compensated for by the control algorithm. This can be achieved by using simple and robust parallel connected on/off valves which operation can be modeled exactly. This enables both fault identification based on an exact machine model and the compensation for the fault by searching for alternative control states.

\subsection{Fault Detection Procedures}

It is crucial that machinery faults can be identified, either by online diagnostics during the normal operation of the machine or by interrupting the machine to run a diagnostics procedure. An approach for identifying faulty digital hydraulics components by a diagnostics procedure is presented in ~\cite{siivonen2007,siivonen2007b}.
Siivonen et al.~\cite{siivonen2007} consider five different cases where a valve may be jammed in either open, closed, or intermediate position, or where the valve does not open or close completely.  
Compared to this method, the approach we present here analyses the machine performance during normal operation of the machine and provides a probabilistic answer to whether any valve is faulty in either its open or closed position. Clearly, the quality of the online fault identification of a valve depends on the activity of that valve during the analyzed time frame. 

\begin{figure}[t]
\centering
\includegraphics[width=0.65\linewidth, trim={0 13.5cm 25.5cm 0},clip]{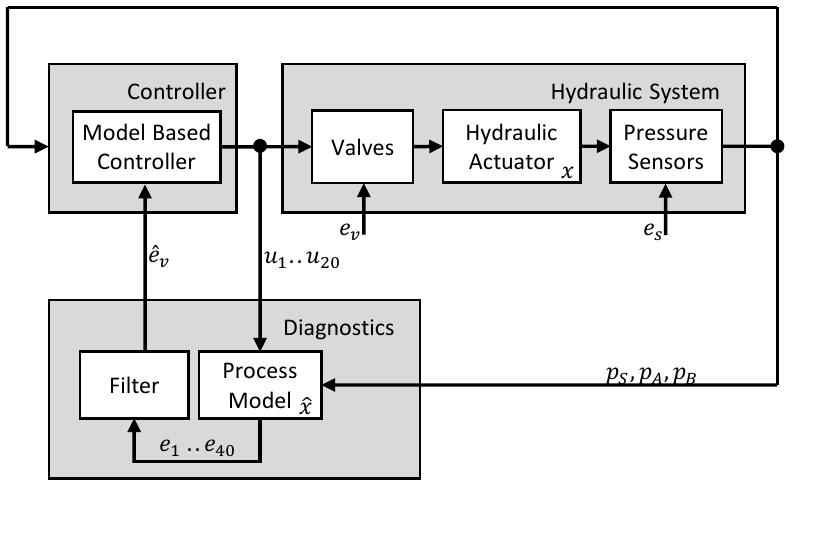}
\caption{The interaction between controller, hydraulic system and online diagnostics. The diagnostics uses the three available pressure sensors and the control signal to estimate the performance of the hydraulic actuator in Figure~\ref{fig:cylinder}.}
\label{fig:model}
\end{figure}

The main components of our approach are presented in Figure~\ref{fig:model}. The hydraulic system has as its input signal the desired state of each of the on/off valves ($u_1 .. u_{20}$) while its outputs are the pressures measured at the hydraulic supply pump $p_S$ and the two cylinder chambers $p_A$ and $p_B$ (compare Figure~\ref{fig:cylinder}). The pressure difference in the cylinder's chambers results in the movement of the piston with velocity $v$, which also depends on the external force $F$. The state of the hydraulic system $x$ represents the actual performance of the hydraulic valves while the state $\hat{x}$ represents the valve performance predicted by the model of the system. This prediction is based on a steady state model that describes a balance of the hydraulic flows in the system, where a broken valve will cause an imbalance which is identifiable by finding the fault that minimizes the difference between the model and the measured behavior of the system.

\subsection{Model-Based and Data-Driven Fault Diagnosis}

Fault-detection techniques are typically based on monitoring measurable variables of a process, generating alarms if a limit-value is reached. Based on the measured variables, fault diagnosis techniques are then used to identify the fault and to take countermeasures. Model-based techniques provide the fault diagnosis with more in-depth information of the process by describing dependencies between different measurable variables~\cite{Isermann200571}. 

Model-based fault diagnosis methods are based on having a mathematical model of the monitored process. The model is usually constructed from physical principles or by system identification~\cite{7069265}. When the process parameters are known, state or output observers can be used to produce an estimate of the real system's state or output respectively~\cite{Isermann200571}. The difference between the outputs of the physical system and the expected outputs, i.e. the residual, is monitored by diagnosis algorithms to detect symptoms of a fault. To isolate several different faults, a bank of observers can be used where a single residual is designed to be sensitive to a specific fault while robust against other faults~\cite{7069265, 5282515}. 

In our case, using observers could mean that we monitor the pressures $p_A$ and $p_B$ and detect a fault in one or more of the valves by comparing the estimated pressures to measured pressures. This would enable us to detect but not to identify a valve fault. Instead our goal is to model each of the hydraulic flows in the system and estimate how each of these match the measurements. The challenge here is that we only measure three pressure values while we estimate 40 variables.

An alternative to model-based fault detection based on observers is parameter estimation~\cite{Isermann200571}. In this approach, some parameters of the process are estimated online and these estimates are compared to the expected parameters of the process~\cite{7069265}. 
Our approach can be regarded as a parameter estimation fault diagnosis method with a large number of parameters to be estimated. As the number of parameters to be estimated is significantly larger than the number of sensors it is necessary to estimate the parameters based on data from a time frame rather than a single time instance. The estimation is an optimization problem where the objective is to minimize the residual for the collected data. The estimated variables will then describe the most likely fault of the system. This focus on data makes our approach resemble data-driven fault diagnosis methods.

Data-driven fault detection methods as opposed to model-driven methods do not typically model the process they monitor but base the analysis on data collected from the process. These methods can be divided into supervised and unsupervised. Supervised methods train a classifier with historical data for both normal and faulty conditions. Representatives of methods that have been used in supervised data-driven fault detection are Support Vector Machines (SVM)~\cite{Yin2016643, Chiang20041389} and Neural  Networks~\cite{1495411}. 
With unsupervised methods, the normal operation of a process is modeled while faults are detected from deviations from the normal operation.  Unsupervised  fault detection methods are threshold based and identify some key features in the analyzed data where a deviation from the normal value indicate a fault. Some important methods are Principal Component Analysis (PCA)~\cite{KU1995179, MACGREGOR1995403},  Independent  Component  Analysis  (ICA)~\cite{Lee20042995}  and  Partial Least  Squares~\cite{MACGREGOR1995403}. 
A comparison of supervised and unsupervised data-driven methods and their ability to detect faults that is not present in the training data is presented in~\cite{6420042}.

Our method is data-driven in the sense that it performs fault identification on measurements from a time frame of one to a few seconds, and identifies the fault that matches the data. Compared to the other mentioned data-driven approaches, our system model is not constructed from data but is constructed as in model-based fault diagnosis techniques. The model combined with the measured data forms the optimization problem describing likely faults. This optimization problem resembles problems from sparse optimization and compressive sensing, and, as we will show later in this paper, similar methods produce a sufficiently precise estimate of the faults in our case.

\subsection{Compressive Sensing}

Measuring the three pressure sensors during the operation of the machine enables us to approximate the flow through each of the 20 valves indicated with $q$ in Figure~\ref{fig:cylinder}. The pressure is thus measured for a number of valve combinations which gives different pieces of information for us to determine the state of the system. This makes our approach related to compressive sensing (or compressive sampling)~\cite{candes2006compressive, Donoho:2006:CS}.

Compressive sensing makes use of the fact that sparse signals can be recovered from far less samples than what is stated by the Shannon-Nyquist sampling theorem~\cite{unser2000, 1580791}. The original signal can then be reconstructed by using $l_1$ optimization to find solutions to an underdetermined linear equation system. Such techniques has been used for e.g. reconstructing images for applications such as magnetic resonance imaging~\cite{lustigMRI} and single pixel cameras~\cite{singelpixel,inpGrLuBj15a}. 

The problem statement for online diagnostics is similar to that of compressive sensing, in that we look for a likely explanation to few observations. Similar to compressive sensing, we also use $l_1$ optimization to solve an linear equation system which is based on measurements, however, in our case the equation system is constructed based on the combination of pressure measurements and the steady state model of the hydraulic system. Another difference is that the goal of compressive sensing is to enable the reconstruction of a signal from a small number of samples while our goal is to identify a fault from a few sensors where the number of samples is not required to be small.

\section{Steady State Model}

The model based diagnostics approach is closely related to model based control of the same system. As the steady state equations are more intuitive in relation to controlling the system, we will start by describing the model based control approach. The control system's task is to choose a control signal that makes the hydraulic actuator follow a desired trajectory as closely as possible while simultaneously minimizing quantities such as energy consumption and valve switching. As an example, a digital valve system may have a number of valve combinations resulting in almost identical movements of the piston but with some combinations having worse energy efficiency due to short circuits such as opening valves on both DFCU PA and DFCU AT simultaneously (see Figure~\ref{fig:cylinder}).

The efficiency of a control signal for a hydraulic system depends on several objectives where the desired trade-off can be described with a penalty function. Finding the optimal control state includes evaluating the steady state achieved by different valve state combinations and finding the best combination with respect to the penalty function. The hydraulic system is described by the steady-state in~\eqref{eq:steadystate}, where the first two equations describe that the difference in flow in and out of a chamber equals the change in the volume of the chamber (movement of the piston). The third equation describes the balance between the external force $F$ and the pressures in the cylinder chambers.

\begin{equation}
\begin{split}
&Q_{PA} - Q_{AT} = A_A \cdot v\\
&Q_{PB} - Q_{BT} = -A_B \cdot v\\
&F = A_A \cdot p_A - A_B \cdot p_B 
\end{split}
\label{eq:steadystate}
\end{equation}

In these equations, the flows $Q$ are given for the path between the supply pump, the cylinder chambers, and the tank. As an example, the flow from the pump to chamber A is given by:
\begin{equation*}
Q_{PA} = \sum_{i=1}^{5} u_{i} \cdot K_{v,i} \cdot SP\left( p_{S} - p_{A} \right)^{\alpha_i}
\end{equation*}
where $u_i$ denotes the 0/1 state of the $i$-th valve, $\alpha$ and $K_v$ are valve parameters and $SP(x)^\alpha$ is the signed power. For a precise model, each valve has individual parameters for $\alpha$ and $K_v$. 

The unknown variables in the steady state equations are the piston velocity $v$ and the pressures in the two cylinder chambers $p_A$ and $p_B$ for the new state. The pressures $p_A$ and and $p_B$ are measured for hte current state and used to compute the external force $F$ but are computed for the prospective next state. The controller's task is hence to find the valve state such that the velocity is close to the desired velocity, the pressures $p_A$ and $p_B$ are within some given limits while minimizing other undesired properties. The steady state describes a balance in the pressures and the flows through the different flow paths.

The steady state in \eqref{eq:steadystate} is not known to have an analytical solution. Instead we need to use numerical methods to find a solution~\cite{linjamadfp16}. As an example, a controller evaluating all $2^{20}$ control states and choosing the optimal one in real-time using Newton-Raphson is presented in~\cite{jedfp16}. By using the model of the system to predict the effect of a specific valve state, the controller can then choose a state that have desired properties. 

The balance we are interested in from a diagnostic point of view, is the balance between the pressure and the hydraulic flows, which can be described as 
\begin{equation}
\frac{Q_{PA} - Q_{AT}}{A_A} + \frac{Q_{PB} - Q_{BT}}{A_B} = 0
\label{eq:scalar}
\end{equation}
where the velocity and the external force have been eliminated from Equation~\ref{eq:steadystate} and we are left with the balance between the flows and the pressures. This means that there is a balance between valve states and the pressures $p_S$, $p_A$ and $p_B$. The pressure difference causes the piston to move, which increases the volume of one cylinder chamber as the other decreases. Equation (\ref{eq:scalar}) simply says that the inflow should be proportional to the outflow, and with the measured pressures the flows through the valves can be predicted.

\section{Fault Estimation}

For fault-tolerance, a machine should be able to identify malfunctioning components. Faulty valves and sensors make the machine work differently than what is predicted by the model based controller, and by estimating these errors a faulty valve can be identified and compensated for. In this section we evaluate a model based error estimation where a sequence of measurements from a few sensors are used to estimate how valves deviate from their expected discrete on/off states. The idea is that this information is collected over every control period (e.g. every 10 ms) while the actual computation of the error estimate can run at a slower rate (e.g. once every second).

\subsection{Deviation between model and measurements}

Starting from Equation~\eqref{eq:scalar} where the unknown variables are $p_A$ and $p_B$, we can revert the problem to known pressures and unknown valve states. With the measured pressures and the corresponding valve states, Equation~\eqref{eq:scalar} should ideally hold. In practice, however, we can expect a small residual $r_{err}$ which represents the difference between the model and the actual behavior of the machine.
\begin{equation}
\frac{Q_{PA} - Q_{AT}}{A_A} + \frac{Q_{PB} - Q_{BT}}{A_B} = r_{err}
\label{eq:residual}
\end{equation}

A faulty valve will cause a considerable residual when the valve is in the faulty state (e.g. the valve is set to be open but is actually closed). To make the model match the actual system better, we evaluate how to introduce a deviation in the discrete control states of the valves, such that the residual is minimized. To make this more general, we evaluate both how a valve deviates from its closed state and its open state.

To start with, we consider only a single time instant and a single evaluation of the balance equation.
If we split Equation~\eqref{eq:residual} into flows through individual valves, the flow through a single open valves of $PA$ can be described as:
\begin{equation*}
q_{PA,i} = u_i \cdot K_{v,i} \cdot SP\left( p_{S} - p_{A} \right)^{\alpha_i}
\end{equation*}
where the variable $u_i$ is $1$ when the valve is set to be open. When the valve is closed the flow should equal $0$, but to enable us to evaluate the error of a valve that does not close properly introduce the complement flow:
\begin{equation*}
q^{c}_{PA,i} = (1 - u_i) \cdot K_{v,i} \cdot SP\left( p_{S} - p_{A} \right)^{\alpha_i}
\end{equation*}
which describes the valve when it is set to be closed. As the control signal to the valve $u_i$ is a binary variable, only one of these two equations will have a nonzero value. With these two variables we can model the error for both a closed and an open valve.

The hydraulic flow through one DFCU can be described as the ideal flow through the valve and a deviation from the two possible states of the valve. We add two variables for estimating the error of each valve,  $x^o$ for a valve which is open and $x^c$ for a closed valve, which should have the values $0$ when the valve is working properly and approach $1$ when the valve is jammed in the corresponding state. The flow through a DFCU with $N$ valves can then be described as:
\begin{equation}
\begin{split}
Q'_{PA} &= \sum_{i=1}^{N} \Big( q_{PA,i} - q_{PA,i}\cdot x^o_i  + q^c_{PA,i} \cdot x^c_i \Big)  \\
  &= Q_{PA} - Q^o_{PA} + Q^c_{PA} \label{eq:onerow}
\end{split}
\end{equation}

Here the first term describes the expected behavior of the valve while the second and third describe the possible faults of the valve.
The variables $x^o$ only have an impact on the equation when the particular valve is open and $x^c$ when the valve is closed. 

\subsection{Describing the most likely fault}

We can describe the steady state equation with prediction of errors by substituting $Q_{PA}$ in Equation~\eqref{eq:scalar} with $Q'_{PA}$ from Equation~\eqref{eq:onerow}, performing similar substitutions for $Q_{AT}$, $Q_{PB}$, and $Q_{BT}$. We can further reorganize the equation such that we can substitute the terms $Q_{PA}$, $Q_{AT}$, $Q_{PB}$, and $Q_{BT}$ with Equation~\eqref{eq:residual} giving us the following equation:

\begin{equation}
\frac{Q^o_{PA}}{A_A} - \frac{-Q^c_{PA}}{A_A} + ... +\frac{Q^c_{BT}}{A_B} = r_{err}
\label{eq:newresid}
\end{equation}

To approximate the source of the residual we need to solve Equation~\eqref{eq:newresid} and thereby approximate the size of the error variables $x^o$ and $x^c$. With the $40$ variables we get from modeling 20 valves, there is an arbitrary number of solutions for the error. However, with many samples, say $K=100$, we can find a solution which minimizes the error for all the samples.

For the whole system, we write the equation system in matrix form $Ax=b$, where each row of \emph{A} corresponds to one sample of Equation~\eqref{eq:newresid} with the corresponding pressure measurements and valve states. 

\begin{equation}
\begin{bmatrix}
 q_{1,1} & q_{1,2} & \cdots & q_{1,40}\\
 q_{2,1} & q_{2,2} & \cdots & q_{2,40}\\
 \vdots & \vdots & \ddots &\vdots \\
 q_{k,1} & q_{k,2} & \cdots & q_{k,40} \\ 
 \cmidrule(lr){1-4}
 p_1 & 0 & \cdots & 0\\
 0 & p_2 & \cdots & 0\\
 \vdots & \vdots & \ddots &\vdots \\
 0 & 0 & \cdots & p_{40}\\
\end{bmatrix}
\begin{bmatrix}
x^o_1\\
x^o_2\\
\vdots\\
x^o_{20}\\
x^c_1\\
x^c_2\\
\vdots\\
x^c_{20}\\
\end{bmatrix}
=
\begin{bmatrix}
r_{err}\\
r_{err}\\
\vdots\\
r_{err}\\
\cmidrule(lr){1-1}
0\\
0\\
\vdots\\
0\\
\end{bmatrix}
\label{eq:leastsquares}
\end{equation}

In Equation~\eqref{eq:leastsquares}, $q_{k,n}$ indicates one flow parameter of the $k$-th sample. The number of unknown variables are 40 as we may have faults both on open and closed valves, and as a consequence the $q_{m,n}$ is given by the potential flow through the respective valves with the measured pressures according to the left hand side terms of Equation~\eqref{eq:newresid} as follows:

\begin{equation*}
q_{k,n} = 
\left\{
\begin{array}{ll}
q_{PA,i} / A_A & \mbox{for } x^o_1 .. x^o_5 \\
-q_{BT,i} / A_B & \mbox{for } x^o_6 .. x^o_{10} \\
-q_{AT,i} / A_A & \mbox{for } x^o_{11} .. x^o_{15} \\
q_{PB,i} / A_B & \mbox{for } x^o_{16} .. x^o_{20} \\
-q^c_{PA,i} / A_A & \mbox{for } x^c_1 .. x^c_5 \\
q^c_{BT,i} / A_B & \mbox{for } x^c_6 .. x^c_{10} \\
q^c_{AT,i} / A_A & \mbox{for } x^c_{11} .. x^c_{15} \\
-q^c_{PB,i} / A_B & \mbox{for } x^c_{16} .. x^c_{20} \\
\end{array}
\right.
\end{equation*}

Ideally, with more samples than unknowns, we get an overdetermined equation system with the unknown vector $\vec{x}$ which describes the error of each of the valves. In practice, it may be underdetermined as the elements in the sensing matrix $A_{m,n}$ is a result of the control signal to the machine and many of its rows may be identical or linear combinations of each other implying that $rank(A) < min(m, n)$. To determine the performance of the valves from Equation~\eqref{eq:leastsquares}, we can use regression to find the errors that best explain the observations.
 
To enforce a solution where the errors are as close as possible to \emph{no error}, we add the lower part of the matrix in Equation~\eqref{eq:leastsquares}, introducing a penalty for leaving the \emph{no error} area. Here the parameters $i_n$ and $p_n$ are tuned such that an error gives a clear offset while no error keeps the error variables close to $0$. In our particular case $p_1$ to $p_{20}$ correspond to average values of the corresponding flows $q$, while $p_{21}$ to $p_{40}$ are $1$.


\begin{figure}[t]
\centering
\includegraphics[width=0.8\linewidth, trim={0cm 0cm 6cm 0cm},clip]{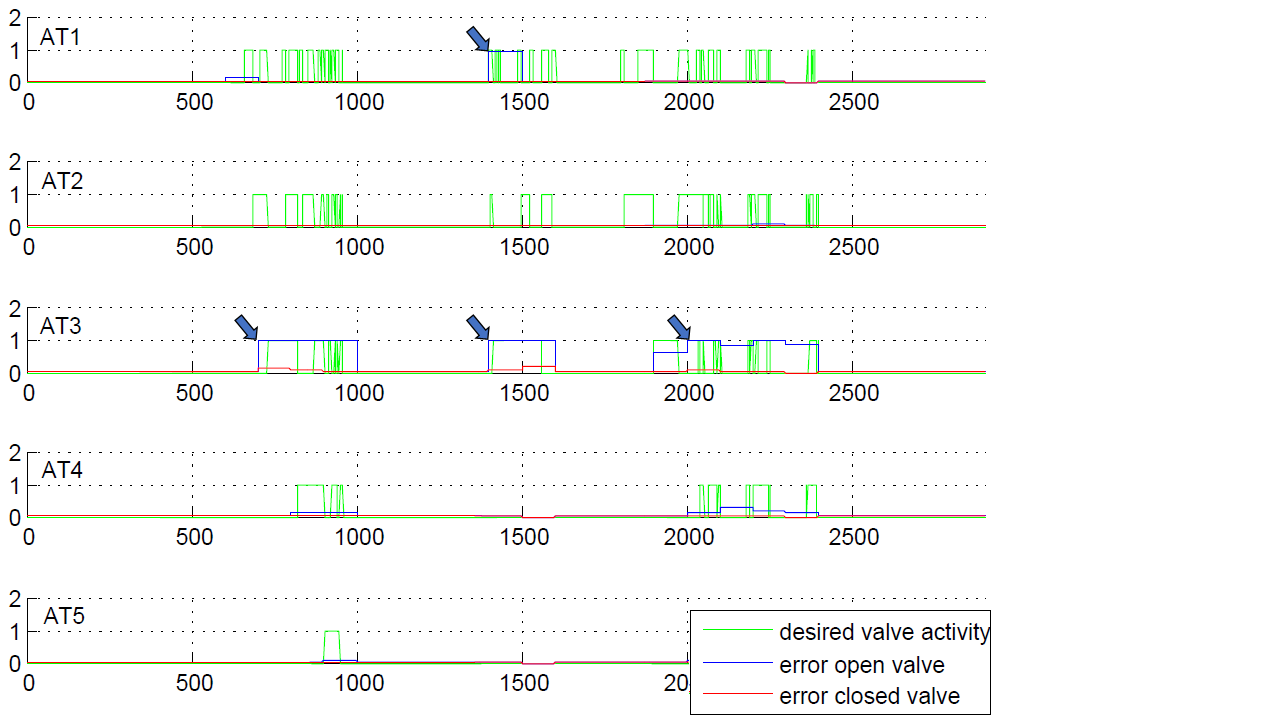}
\caption{Example of Quantile Regression where the valve AT3 is jammed in closed position. It can be seen that \emph{error open} is close to one (i.e. indicating fault) during the periods when the valve is actively used. Also AT1 indicates a fault for a short period, however, it can be statistically identified not to have significance.}
\label{fig:ex3}
\end{figure}

\subsection{Fault identification by Regression}

A solution to Equation~\eqref{eq:leastsquares} approximates how much the valves deviate from the discrete on/off setting. An example of the result with one jammed valve is shown in Figure~\ref{fig:ex3}, where the linear equation system is solved for time frames of one second by Quantile Regression (QR)~\cite{QR2001}. The results in Figure~\ref{fig:ex3} show the obtained predictions of the faults for each of the valves in DFCU $AT$ as well as the valve state of each of these valves. In this example, the cable between the valve \emph{AT3} and the controller has been physically unplugged which results in the valve being jammed in its closed state. We can see that during intervals where valve \emph{AT3} is set to open, the \emph{error open} is close to $1$. 

To obtain the solution for Equation~\eqref{eq:leastsquares}, we want to minimize the absolute deviations (i.e. $l_1$ regression) meaning that we compute the median of the (conditional) distribution. Finding the median is a special case of QR, where QR is defined by
\[
\underset{x_i \in 0..1}{\min} \rho_{\tau} (b-Ax)
\]
where
\[
\rho_{\tau}(u) = \sum u_i  \cdot 
\left\{
\begin{array}{ll}
\tau,  &  u_i \geq 0\\
\tau - 1,  & u_i<0
\end{array}
\right.
\]
giving the weighted sum of absolute residuals, which in our case simply is the median as we use the parameter $\tau = 0.5$. This minimization problem can be solved efficiently by linear programming~\cite{QR2001,YangMM}.

While we obtain good results with this approach, we observe that also other valves occasionally show deviation, e.g. due to noise in the pressure measurements. For this reason it is necessary to collect more information about each valve and, based on the history, approximate the error.

\begin{figure}[t]
\centering
\includegraphics[width=0.7\linewidth, trim={4cm 11.5cm 4cm 11cm},clip]{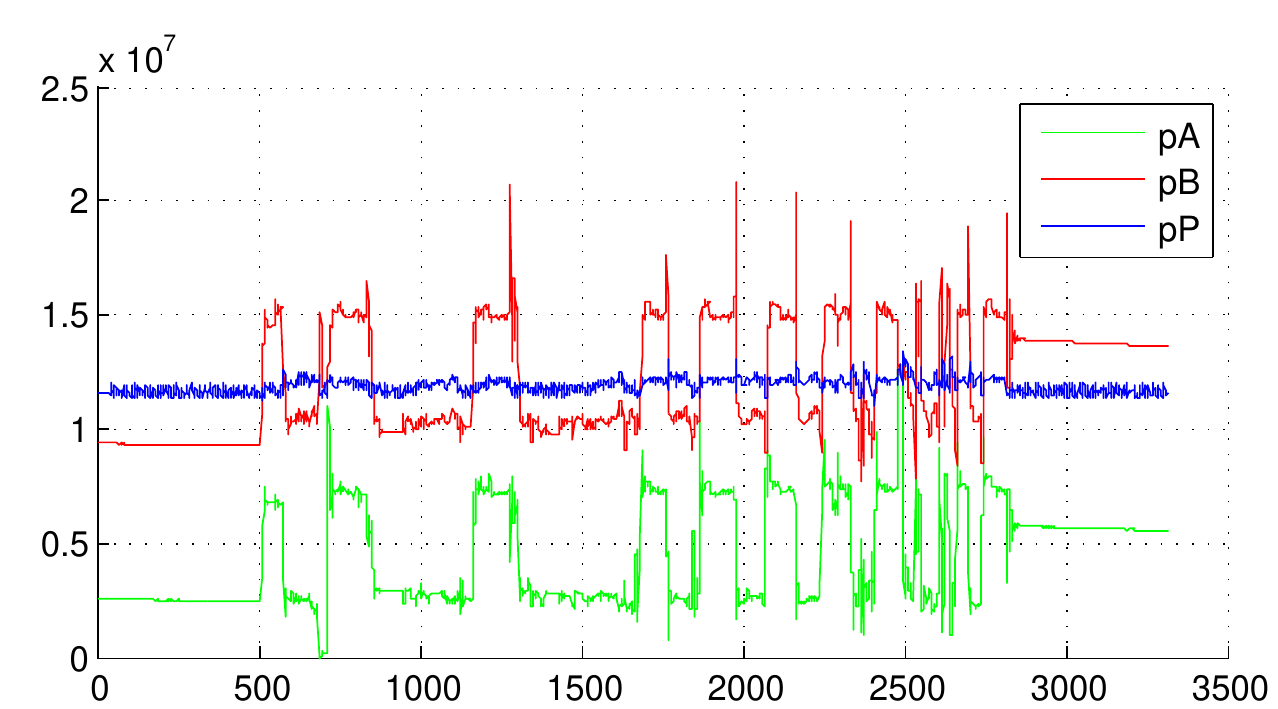}
\caption{The measured pressures showing noise and both plateaus and spikes in the chamber pressures. The spikes cause false positives in the fault identification.}
\label{fig:spikes}
\end{figure}

\section{Experimental Validation}

There are various scenarios where the identification described in the previous section will cause false positives. Already by studying Figure~\ref{fig:ex3} we can see a disturbance at around sample 1500 on the valve AT1 which is caused by a fast change of piston direction which causes oscillations in the measured pressures. During these circumstances, the steady state model does not properly describe the behavior of the physical system. This means that the estimated error is the error between the physical system and the steady state model instead of a faulty valve. For this reason it is important not to draw conclusions based on a single deviation between the model and the measurements. 

Another important observation, which also is visible in Figure~\ref{fig:ex3}, is that actual faults are constantly present when there is activity of the broken valve while false positives seems to occur more sporadically (e.g. see AT1 at 1500 ms in Figure~\ref{fig:ex3}). In practice, we know that the false positives often are a result of fast changes in pressure. By combining this information, we can eliminate many of the false positives by removing samples around pressure peaks (see Figure~\ref{fig:spikes}) and by estimating faults based on evidence collected during a longer time period.

The results from solving the equation system in Equation~\eqref{eq:leastsquares} gives the predicted state of the valves for periods of 100 samples (which corresponds to one second in our case). To make the prediction more robust against noise and thus avoiding false positives, the diagnostics should not be based on single periods but instead slowly evolve during many periods. For this purpose we filter the fault approximation of each valve such that the impact of old data receive less weight with time. 

In practice we implement a filter that computes the Exponentially Weighted Moving Average (EWMA)~\cite{nisthandbook}. This is simply an infinite impulse response (IIR) filter 
\[y_k = \lambda \cdot u_k + (1-\lambda) \cdot y_{k-1}\]
where $y$ is the prediction, $u$ is a new measurement, and $\lambda$ is the weighting factor which determines how fast the filter reacts to new data. To get an accurate prediction of the valve operation, a new prediction of the specific valve is only computed from periods where the estimated property of the valve has been active. As an example, if we estimate the error of a valve in open position, the evidence collected during a period where the valve never was in open position, is not useful. For this reason, the estimate of a single valve fault is only used from periods where the proportion of samples with evidence of that fault reaches a threshold (10\% in the presented results).

\begin{table}[]
\renewcommand{\arraystretch}{1.2}
\centering
\caption{Comparison of identified faults and the largest fault shown for valves working properly. Values close to 1 identifies a fault. Applying a filter effectively removes the problem of false positives. The asterisk means synthetic data.}
\label{tab:cases}
\begin{tabular}{llllll}

\hline
\multicolumn{1}{|l}{Q-Regress}  & \multicolumn{2}{c}{Filtered} & \multicolumn{2}{c|}{Not filtered}\\ \hline 
\multicolumn{1}{|l|}{Scenario}  & \multicolumn{1}{l}{Fault}  & \multicolumn{1}{l|}{False} &
\multicolumn{1}{l}{Fault}  & \multicolumn{1}{l|}{False} \\ \hline

\multicolumn{1}{|l|}{No fault 1}  & \multicolumn{1}{c}{-} & \multicolumn{1}{c|}{ 0.12} & 
\multicolumn{1}{c}{-} & \multicolumn{1}{c|}{0.37 } \\ 
\multicolumn{1}{|l|}{No fault 2}  & \multicolumn{1}{c}{-} & \multicolumn{1}{c|}{ 0.08}  &
\multicolumn{1}{c}{-} & \multicolumn{1}{c|}{0.40 } \\ 
\multicolumn{1}{|l|}{No fault 3}  & \multicolumn{1}{c}{-} & \multicolumn{1}{c|}{0.13}  &
\multicolumn{1}{c}{-} & \multicolumn{1}{c|}{ 0.87} \\ 
\multicolumn{1}{|l|}{No fault 4}  & \multicolumn{1}{c}{-} & \multicolumn{1}{c|}{ 0.11}  &
\multicolumn{1}{c}{-} & \multicolumn{1}{c|}{ 0.76} \\ 
\multicolumn{1}{|l|}{Closed AT1}  & \multicolumn{1}{c}{ 0.40} & \multicolumn{1}{c|}{ 0.14}  &
\multicolumn{1}{c}{ 1.0} & \multicolumn{1}{c|}{ 0.71} \\ 
\multicolumn{1}{|l|}{Closed AT3}  & \multicolumn{1}{c}{ 0.92} & \multicolumn{1}{c|}{ 0.15}  &
\multicolumn{1}{c}{ 1.0} & \multicolumn{1}{c|}{ 1.0} \\ 
\multicolumn{1}{|l|}{Closed AT5}  & \multicolumn{1}{c}{ 0.96} & \multicolumn{1}{c|}{ 0.33}  &
\multicolumn{1}{c}{ 1.0} & \multicolumn{1}{c|}{ 1.0} \\  
\multicolumn{1}{|l|}{Closed PA3}  & \multicolumn{1}{c}{ 0.91} & \multicolumn{1}{c|}{ 0.18}  &
\multicolumn{1}{c}{1.0 } & \multicolumn{1}{c|}{ 1.0} \\  
\multicolumn{1}{|l|}{Closed PB3}  & \multicolumn{1}{c}{ 0.90} & \multicolumn{1}{c|}{ 0.18}  &
\multicolumn{1}{c}{1.0 } & \multicolumn{1}{c|}{1.0 } \\ 
\multicolumn{1}{|l|}{Open PB3*}  & \multicolumn{1}{c}{0.91} & \multicolumn{1}{c|}{0.10}  &
\multicolumn{1}{c}{0.97} & \multicolumn{1}{c|}{0.22} \\ \hline 

\end{tabular}
\end{table}

The results in Table~\ref{tab:cases} show the impact of using a filter with $\lambda = 0.10$ (\emph{Filtered}) compared to the largest errors shown in the input to the filter (\emph{Not filtered}). The experiment includes four sample runs of a hydraulic excavator boom where all valves work properly and five runs where one valve has been jammed in closed position by physically unplugging the control signal. In the table, the columns named \emph{Fault} represent the estimation of the actual fault while the columns named \emph{False} represent the larges fault shown on a properly working valve. As we can see, the filter effectively removes false positives when the system is working properly while letting the actual faults slowly appear when present. 

For completeness, a single experiment based on synthetic data for a valve jammed in open position was added to verify that the approach can identify both types of faults. The synthetic data for \emph{Open PB3} was created from the parts of the measured sequences where \emph{PB3} was in the open state while the control signal to this valve was a random sequence. This means that the precision of the fault identification in \emph{Open PB3} is not interesting, instead it only shows that the approach can distinguish between both types of faults.

The results presented in Table~\ref{tab:cases} are based om measurements collected from an excavator boom and analyzed with Matlab. The used measurements including measured pressures of the excavator boom with the corresponding valve states and control signals are available at~\cite{data}. The corresponding functionality was also implemented on NVIDIA Jetson-tk1~\cite{jetson}, which was connected to the excavator boom's control system through a CAN-bus. A demonstration video of running the fault identification during normal operation of the machine while a valve becomes broken (by being unplugged) can be viewed at~\cite{thevideo}.

\section{Conclusion}
In this paper we present an approach to identify faults in the operation of a large number of hydraulic on/off valves based on sampling a small number of pressure sensors during the normal operation of a machine. The approach does not require adding new sensors to the hydraulic system, but only uses sensors that are needed for controlling the hydraulic actuator. A steady state model of the hydraulic system is used to estimate the hydraulic flow through the valves, based on the measured pressures. When the difference between the hydraulic system  and the model is large, the most probable cause for the deviation is computed using quantile regression.  

The experiments, performed on an excavator boom, show that jammed valves can be precisely identified almost in real-time while the machine operates. This enables more efficient service as the faulty component is automatically identified potentially before the operation notices the fault. It also enables utilizing the redundancy of digital hydraulic systems, enabling a machine to be fully operational despite a faulty valve, as a controller which is aware of a fault can choose to control the hydraulic actuator with control states which are not affected by the fault.  

\section*{Acknowledgements}
This work has been funded by the Academy of Finland project Merge (No. 286094).

\section*{References}

\bibliographystyle{plain}
\bibliography{references}

\begin{thebibliography}{10}

\bibitem{thevideo}
Online diagnostic for digital hydraulics.
\newblock Demonstration video, {Y}outube 2017.

\bibitem{blanke:2006}
Mogens Blanke, Michel Kinnaert, Jan Lunze, Marcel Staroswiecki, and
  J.~Schr\"{o}der.
\newblock {\em Diagnosis and Fault-Tolerant Control}.
\newblock Springer-Verlag New York, Inc., Secaucus, NJ, USA, 2006.

\bibitem{1580791}
E.~J. Candes, J.~Romberg, and T.~Tao.
\newblock Robust uncertainty principles: exact signal reconstruction from
  highly incomplete frequency information.
\newblock {\em IEEE Transactions on Information Theory}, 52(2):489--509, Feb
  2006.

\bibitem{candes2006compressive}
Emmanuel~J Cand{\`e}s et~al.
\newblock Compressive sampling.
\newblock In {\em Proceedings of the international congress of mathematicians},
  volume~3, pages 1433--1452. Madrid, Spain, 2006.

\bibitem{Chiang20041389}
Leo~H. Chiang, Mark~E. Kotanchek, and Arthur~K. Kordon.
\newblock Fault diagnosis based on fisher discriminant analysis and support
  vector machines.
\newblock {\em Computers \& Chemical Engineering}, 28(8):1389 -- 1401, 2004.

\bibitem{Donoho:2006:CS}
D.~L. Donoho.
\newblock Compressed sensing.
\newblock {\em IEEE Trans. Inf. Theor.}, 52(4):1289--1306, April 2006.

\bibitem{singelpixel}
M.~F. Duarte, M.~A. Davenport, D.~Takbar, J.~N. Laska, T.~Sun, K.~F. Kelly, and
  R.~G. Baraniuk.
\newblock Single-pixel imaging via compressive sampling.
\newblock {\em IEEE Signal Processing Magazine}, 25(2):83--91, March 2008.

\bibitem{jedfp16}
J~Ersfolk, P~Boström, V~Timonen, J~Westerholm, J~Wiik, O~Karhu, M~Linjama, and
  M~Waldén.
\newblock Optimal digital valve control using embedded gpu.
\newblock In {\em Proceedings of the Eight Workshop on Digital Fluid Power},
  pages 239 -- 250, 2016.

\bibitem{7069265}
Z.~Gao, C.~Cecati, and S.~X. Ding.
\newblock A survey of fault diagnosis and fault-tolerant techniques -- part i:
  Fault diagnosis with model-based and signal-based approaches.
\newblock {\em IEEE Transactions on Industrial Electronics}, 62(6):3757--3767,
  June 2015.

\bibitem{6420042}
M.~Grbovic, W.~Li, N.~A. Subrahmanya, A.~K. Usadi, and S.~Vucetic.
\newblock Cold start approach for data-driven fault detection.
\newblock {\em IEEE Transactions on Industrial Informatics}, 9(4):2264--2273,
  Nov 2013.

\bibitem{inpGrLuBj15a}
Stefan Grönroos, Wictor Lund, and Jerker Björkqvist.
\newblock A single pixel camera based on a dlp video projector.
\newblock In Sirkka-Liisa Jämsä-Jounela, editor, {\em Proceedings of
  Automaatio XXI}, volume~44, page 1–6. Suomen Automaatioseura ry Finnish
  Society of Automation, 2015.

\bibitem{5282515}
I.~Hwang, S.~Kim, Y.~Kim, and C.~E. Seah.
\newblock A survey of fault detection, isolation, and reconfiguration methods.
\newblock {\em IEEE Transactions on Control Systems Technology},
  18(3):636--653, May 2010.

\bibitem{Isermann200571}
Rolf Isermann.
\newblock Model-based fault-detection and diagnosis – status and
  applications.
\newblock {\em Annual Reviews in Control}, 29(1):71 -- 85, 2005.

\bibitem{QR2001}
Roger Koenker and Kevin~F. Hallock.
\newblock Quantile regression.
\newblock {\em The Journal of Economic Perspectives}, 15(4):143--156, 2001.

\bibitem{KU1995179}
Wenfu Ku, Robert~H. Storer, and Christos Georgakis.
\newblock Disturbance detection and isolation by dynamic principal component
  analysis.
\newblock {\em Chemometrics and Intelligent Laboratory Systems}, 30(1):179 --
  196, 1995.

\bibitem{Lee20042995}
Jong-Min Lee, ChangKyoo Yoo, and In-Beum Lee.
\newblock Statistical monitoring of dynamic processes based on dynamic
  independent component analysis.
\newblock {\em Chemical Engineering Science}, 59(14):2995 -- 3006, 2004.

\bibitem{linjamadfp16}
M~Linjama.
\newblock On the numerical solution of steady-state equations of digital
  hydraulic valve-actuator system.
\newblock In {\em Proceedings of the Eight Workshop on Digital Fluid Power},
  pages 141 -- 155, 2016.

\bibitem{Linjama:2011}
Matti Linjama.
\newblock Digital fluid power – state of the art.
\newblock In {\em Proceedings of the Twelfth Scandinavian International
  Conference on Fluid Power, SICFP'11}, 2011.

\bibitem{data}
Matti Linjama, Miika Ahopelto, and Johan Ersfolk.
\newblock Digital hydraulic boom with valve faults.
\newblock Etsin research data finder, 2017-05-08.

\bibitem{linjama2007}
Matti Linjama, Mikko Huova, Pontus Boström, Arto Laamanen, Lauri Siivonen,
  Lionel Morel, Marina Waldén, and Matti Vilenius.
\newblock Design and implementation of energy saving digital hydraulic control
  system.
\newblock In J.~Vilenius, K.~T. Koskimies, and J.~Uusi-Heikkilä, editors, {\em
  Proceedings of 10th Scandinavian International Conference on Fluid Power
  (SICFP'07)}, volume~2, pages 341--359. Tampere University of Technology,
  2007.

\bibitem{lustigMRI}
M.~Lustig, D.~L. Donoho, J.~M. Santos, and J.~M. Pauly.
\newblock Compressed sensing mri.
\newblock {\em IEEE Signal Processing Magazine}, 25(2):72--82, March 2008.

\bibitem{MACGREGOR1995403}
J.F. MacGregor and T.~Kourti.
\newblock Statistical process control of multivariate processes.
\newblock {\em Control Engineering Practice}, 3(3):403 -- 414, 1995.

\bibitem{nisthandbook}
NIST/SEMATECH.
\newblock {EWMA} {C}ontrol {C}harts.
\newblock e-Handbook of Statistical Methods.

\bibitem{jetson}
NVIDIA.
\newblock Jetson-tk1 embedded development platform.
\newblock 2017.

\bibitem{siivonen2007b}
L.~Siivonen, M.~Linjama, M.~Huova, and M.~Vilenius.
\newblock Fault detection and diagnosis of digital hydraulic valve system.
\newblock In {\em The Tenth Scandinavian International Conference on Fluid
  Power, May 21-23, 2007, Tampere, Finland, SICFP\'07}. Tampere University of
  Technology, 2007.

\bibitem{siivonen2007}
L.~Siivonen, M.~Linjama, M.~Huova, and M.~Vilenius.
\newblock Pressure based fault detection and diagnosis of a digital valve
  system.
\newblock In {\em Power Transmission and Motion Control (PTMC 2007), University
  of Bath, UK, 12-14 September 2007}, 2007.

\bibitem{1495411}
S.~Simani.
\newblock Identification and fault diagnosis of a simulated model of an
  industrial gas turbine.
\newblock {\em IEEE Transactions on Industrial Informatics}, 1(3):202--216, Aug
  2005.

\bibitem{unser2000}
M.~Unser.
\newblock Sampling---50 {Y}ears after {S}hannon.
\newblock {\em Proceedings of the {IEEE}}, 88(4):569--587, April 2000.

\bibitem{YangMM}
Jiyan Yang, Xiangrui Meng, and Michael~W. Mahoney.
\newblock {Quantile Regression for Large-scale Applications}.
\newblock In {\em {Proceedings of the 30th International Conference on Machine
  Learning}}, volume~28 of {\em {JMLR Proceedings}}, pages 881--887.
  {JMLR.org}, 2013.

\bibitem{Yin2016643}
Zuyu Yin and Jian Hou.
\newblock Recent advances on \{SVM\} based fault diagnosis and process
  monitoring in complicated industrial processes.
\newblock {\em Neurocomputing}, 174, Part B:643 -- 650, 2016.

\end{thebibliography}

\end{document}